\documentclass[notitlepage,12pt]{article}%

\usepackage{amsfonts}
\usepackage{amssymb}
\usepackage{amsmath}
\usepackage{amsthm}
\usepackage{comment}
\usepackage[normalem]{ulem}
\usepackage{multirow}
\usepackage[flushleft]{threeparttable}
\usepackage{threeparttablex}
\usepackage{array}
\usepackage{setspace}
\usepackage{ragged2e}
\usepackage{lscape}
\usepackage{graphicx}
\graphicspath{ {./images/} }
\usepackage{amsfonts}
\usepackage{scalefnt}
\usepackage{natbib}
\usepackage{hyperref}
\usepackage{caption}
\usepackage{subcaption}
\usepackage{appendix}
\usepackage{wrapfig}

\usepackage{booktabs}
\usepackage{siunitx}
\usepackage[margin=1in]{geometry}
\newcommand{\indep}{\perp \!\!\! \perp}

\theoremstyle{plain}
\newtheorem{theorem}{Theorem}

\theoremstyle{definition}

\newtheorem{example}{Example}
\newtheorem{assumption}{Assumption}
\newtheorem{algorithm}{Algorithm}
\newtheorem{remark}{Remark}

\begin{document}
\thispagestyle{empty}

\title{\fontsize{17.8}{19}\selectfont \bfseries Endogenous Control Bias and Correction with Nonparametric Methods}
\author{
Kaicheng Chen\thanks{School of Economics, Shanghai University of Finance and Economics.
Email: chenkaicheng@sufe.edu.cn}
\and Kyoo il Kim\thanks{Department of Economics, Michigan State
University. Email: kyookim@msu.edu}}
\date{}

\begingroup
\renewcommand{\thefootnote}{}
\footnotetext{This paper supersedes the earlier manuscript ``Identification of Average Responses with Endogenous Controls'' (arXiv:2401.14395).}

\endgroup

\maketitle

\begin{abstract}

Controls are often needed to make a treatment or instrument conditionally valid, yet they may be related to unobserved outcome determinants. We show that restricted parametric adjustment can fail to recover the structural average response even when the structural model is correctly specified. Under conditional independence and sufficient conditional treatment variation, flexible conditional mean methods identify the structural average response in a nonseparable model, while a control function extension accommodates endogenous treatments with conditionally valid instruments. We develop a specification test to guide the choice between restricted and flexible functional forms. Simulations and a China shock application illustrate the methods.\\




\end{abstract}

\setcounter{page}{0} \thispagestyle{empty} \pagestyle{plain}

\newpage

\section{ Introduction}

Empirical researchers routinely add controls to account for confounders. This practice creates a two-sided problem. Omitting relevant controls may leave the treatment or instrument endogenous, but including them in an arbitrary way may not remove its entire relationship with unobserved outcome determinants. The difficulty is especially important when a control is needed for conditional validity but its relationship with the structural disturbance is unrestricted.

We call such a variable $X$ an \emph{endogenous control}: its dependence with the structural disturbance is unrestricted, while conditioning on it renders the treatment conditionally independent of the disturbance or the instrument conditionally valid. Note that this definition does not admit a post-treatment control, or ``bad control'', because, with the presence of bad control, conditional validity is not likely to hold (\citealp{lechner2008note}, \citealp{wooldridge2005violating}). 

In this paper, we show how endogenous controls can induce bias in simple linear and nonlinear models, and illustrate the issue in empirically relevant settings such as shift-share IV designs. Restricted inclusion of \(X\), as in conventional parametric methods—for example, OLS, 2SLS, and probit estimation—may fail to identify the structural average effects of interest. Meanwhile, endogenous controls need not preclude identification; instead, they change how conditioning must be implemented. 
We carefully define the target structural parameter for a broad class of models allowing endogenous covariates (treatments and controls), and show how identification can be restored through an appropriate conditioning argument under conditional validity and sufficient conditional treatment variation. The implied flexible estimation is nonparametric and can be readily implemented through existing statistical packages.

To fix ideas, suppose the structural equation is correctly specified as
\begin{align}
    Y = h(W,X,\varepsilon;\theta) \label{eq:parametric}
\end{align}
where $W$ is the scalar variable of interest, $X$ is a vector of controls, and $h$ is known up to $\theta$. The structural average response of interest is
\begin{equation} \label{eq:sar_def}
    \beta_0 = E[\partial_W h(W,X,\varepsilon;\theta)].
\end{equation}
The structural equation may be correctly specified while the estimand corresponding to OLS, 2SLS, or other restricted nonlinear parametric estimators does not coincide with the structural parameter of interest. If the distribution or conditional mean of $\varepsilon$ varies with $X$, restricted inclusion of $X$ can leave control related components in the estimating error. Endogenous control bias is therefore closely related to misspecification bias. 

Consider a linear model as a simple motivating example:
\begin{align}
Y=\tau W+X'\xi+\varepsilon, \qquad E[\varepsilon\mid W,X]=E[\varepsilon\mid X]. \label{eq:intro_linear}
\end{align}
Here, the structural average response with respect to the treatment $W$ is given by $\tau$. Due to dependence between $X$ and the structural disturbance $\varepsilon$, linear inclusion of $X$ need not recover $\tau$. In contrast, when the conditional mean is differentiable and there is sufficient conditional variation in $W$, we obtain 
\begin{align}
\tau=E[\partial_W E(Y\mid W,X)]. \label{eq:intro_ade}
\end{align}
The same dilemma arises in IV designs: controls may be needed for instrument validity, yet linear residualization may be inadequate. A control function can first restore conditional independence, after which flexible conditional mean estimation recovers the response.

These examples raise broader questions: what structural response a flexible conditional mean method recovers, under what conditions, and what applied researchers can learn from disagreement between restricted and flexible estimates. To answer these questions, we make the following contributions. First, we unify the motivating OLS, IV as well as nonlinear examples through an average response parameter in a general nonseparable model. Second, we provide constructive identification of the average response parameter through an conditioning argument. The parameter of interest as a function of conditional means then can be estimated through standard nonparametric methods. Thirdly, as an implication of the endogenous bias under the parametric specification, we further develop a specification test to guide the choice between the restricted and flexible functional forms. Lastly, simulations and a reexamination of the China shock study \citep{autor2013china} show how the proposed method can materially affect empirical conclusions.

The general result requires both conditional validity and sufficient treatment variation. Conditional independence holds the disturbance distribution fixed as $W$ varies after conditioning. For a continuous treatment, local conditional support and smoothness make the relevant derivative well-defined and identified from observable conditional means; for a discrete treatment, conditional overlap makes the corresponding difference identified. These conditions can hold when controls are endogenous in our sense, but fail when treatment cannot be varied while controls are held fixed. They do not rehabilitate bad controls or compensate for weak or absent conditional treatment variation.

Our analysis connects four literatures. The motivating specification problem is related to work on endogenous controls and flexible adjustment, including \cite{frolich2008parametric}, \cite{il2013regression}, \cite{diegert2022assessing}, and \cite{iv2025}. The average response framework builds on nonseparable identification under conditional independence, including \cite{altonji2005cross} and \cite{imbens2009identification}. The distinction between legitimate conditioning variables and bad controls follows \cite{wooldridge2005violating}, \cite{lechner2008note}, and \cite{angrist2009mostly}. Finally, our conditional IV discussion is related to control function methods and to shift-share designs. We revisit the shift-share design of \cite{autor2013china} to study the China shock.


The remainder of the paper is organized as follows. Section \ref{sec:motivation} develops the motivating examples further and demonstrates the general results in linear models as well as a nonlinear binary model. Section \ref{sec:nonseparable} obtains the main identification results for a broad class of models. Section \ref{sec:test} develops a test based on the difference between restricted parametric and flexible nonparametric estimates of the average response. Section \ref{sec:sim} reports the simulation evidence. Section \ref{sec:example} revisits the China shock study. Section \ref{sec:conclu} concludes the paper with empirical guidance.

\textbf{Notation.} Throughout the paper, we let $F(A,B,C)$, $F(A,B)$, and $F(A)$ denote the distribution functions of $(A,B,C)$, $(A,B)$, and $A$; $F_{A|B,C}$ denotes the distribution functions of $A$ conditional on $B,C$. Whenever a (conditional) density is invoked, we assume it exists and is well-defined. We denote (conditional) density functions $f$ similarly. $A \indep B \ |\ C$ denotes the independence of $A$ and $B$ conditional on $C$. The supports of the random elements $A$ are denoted by their calligraphic letters $\mathcal{A}$. We use $P$ to denote the probability measure and $P^*$ to denote the bootstrap probability measure conditional on the observed data.

\section{Motivating examples}
\label{sec:motivation}

\subsection{Linear regression}
To illustrate the problem, let's consider the following linear model:
\begin{align}
    Y = \tau W + X'\xi + \varepsilon , \ E[\varepsilon|W,X]=E[\varepsilon|X]. \label{eq:linear}
\end{align}
In this model, the structural target is $\beta_0=\tau$ by definition in \eqref{eq:sar_def}. The disturbance is mean independent of $W$ conditional on $X$, but it need not be mean independent of $X$. There are two consequences. Omitting $X$ may generate omitted variable bias, while including $X$ linearly need not eliminate the control related component of the disturbance. Let $D=(W,X')'$ and $\theta=(\tau,\xi')'$. The restricted OLS estimand is
\begin{equation}
        \gamma_{LP}\equiv E[DD']^{-1}E[DY] \nonumber=\theta+E[DD']^{-1}E\left[D E(\varepsilon\mid X)\right]. \label{eq:linear_projection_bias}
\end{equation}
Thus, $\tau$ need not be identified by the linear projection.
We denote the first component of $\gamma_{LP}$ by $\beta_p$; it is the restricted estimand in this example.

Flexible adjustment can nevertheless identify $\tau$ when conditional independence is accompanied by sufficient treatment variation after conditioning. The idea is as follows. Under the parametric specification assumption \eqref{eq:parametric}, there exists a nonparametric and nonseparable nesting model $m(W,X,\varepsilon)$ such that 
$$
m(W,X,\varepsilon) = h(W,X,\varepsilon;\theta) = Y, 
$$
almost surely. It follows that the average response defined from this nesting model is equivalent to the average response defined by $h$, i.e., $
    E[\partial_W m(W,X,\varepsilon)] = E[\partial_W h(W,X,\varepsilon;\theta)] = \beta_0.
$
It is shown later that, if $W$ is independent of $\varepsilon$ given $\tilde{X}$, where the \(\sigma\)-field generated by $X$ satisfies $\sigma(X)\subseteq \sigma(\widetilde X)$ for some $\widetilde X$, then
\begin{align}
    \beta_0 = E[\partial_W m(W,X,\varepsilon)]  =  E[\partial_W E(Y|W,\widetilde{X})] \label{eq:main_result}    
\end{align}
For model \eqref{eq:linear}, take $\widetilde X=X$ and define the flexible estimand $\beta_{np}=E[\partial_W E(Y\mid W,X)]$. Under the maintained conditions,
\begin{align*}
\int_{\mathcal{W}\times\mathcal{X}}\partial_wE[Y|W=w,X=x]dF(w,x)= \tau+\int_{\mathcal{W}\times\mathcal{X}}\partial_w(x'\xi+E[\varepsilon|X=x])dF(w,x)=\tau,
\end{align*}
so $\beta_{np}=\beta_0=\tau$. 

The above relation holds simply under the conditional independence. However, the identification is not yet complete: for example, if $W$ is completely determined by $X$, then the treatment cannot be varied while holding the controls fixed, in which case the derivative of the conditional mean is not identified. Therefore, we need a local support condition so that the conditional derivative is identified from the data. For a continuous $W$, the observed treatment values over which the response is averaged must lie in the interior of the conditional support of $W\mid X=x$, and the conditional mean must be sufficiently smooth in $w$. This local support condition can be viewed as a conditional rank or variation condition: it permits $W$ to vary locally while $X$ is held fixed. It is weaker than rectangular support because the relevant treatment range may differ across values of $X$.

For a binary $W$, differentiation is replaced by a discrete difference and local interior support is replaced by conditional overlap. Specifically, both treatment states must occur with positive conditional probability at almost every value of $X$ over which the response is averaged:
\[
0<P(W=1\mid X=x)<1.
\]
Combining this condition with conditional mean independence gives
\begin{align*}
    \int_{\mathcal X}
    \left\{
    E[Y\mid W=1,X=x]
    -
    E[Y\mid W=0,X=x]
    \right\}
    dF(x)
    =\tau.
\end{align*}

Therefore, two ingredients together deliver the identification of average response of the nesting model, which in turn recovers the structural parameter from the original model. The first ingredient conditional independence connects the structural response to a derivative or difference of observable conditional means. Local support for a continuous treatment, or overlap for a discrete treatment, supplies the within-control treatment variation needed to identify that derivative or difference. Smoothness is additionally required for derivative based identification. Section~\ref{sec:nonseparable} states these conditions formally.

\begin{example}
    As a concrete example, consider the linear regression model relating district level average test scores (\textit{avgscore}) to district level educational expenditure per student (\textit{expend}) and average family income (\textit{avginc}) from \citealp[Chapter 3]{wooldridge2019introductory}:
\begin{align*}
    avgscore = \alpha + \tau \cdot expend + \xi \cdot avginc + \varepsilon.
\end{align*}
Suppose we are interested in the average response associated with \textit{expend}, which is $\tau$ in this case. Since \textit{avginc} is relevant for \textit{expend} at the district level and \textit{avginc} can also affect \textit{avgscore} through other channels, e.g., private tutoring, including \textit{avginc} as a control variable is sensible. However, \textit{avginc} may also be correlated with other unobserved determinants that affect both \textit{avginc} and \textit{avgscore}; the endogeneity of \textit{avginc} can then contaminate estimation of $(\tau,\xi)$ by OLS. Nevertheless, $\tau$ is nonparametrically identified if (1) \textit{expend} is independent of $\varepsilon$ conditional on \textit{avginc}, (2) \textit{expend} has sufficient local variation conditional on \textit{avginc}, and (3) the relevant conditional mean is smooth in \textit{expend}. 
\end{example} 

\subsection{Linear IV model}
Controls are often included to make an instrument valid rather than merely to improve precision. Consider the linear triangular model
\begin{align*}
    Y &= \tau W + X'\xi+\varepsilon, \ E[\varepsilon|W,X]\ne E[\varepsilon|X] \\
    W &= \pi_Z Z + \pi_X X + \eta.
\end{align*}
Here, the average response $\beta_0=\tau$, $W$ remains endogenous conditional on $X$, and $Z$ is maintained to be valid only conditional on $X$. Let $M_X$ denote population linear residualization on $X$. The restricted IV estimand for $\tau$ is
\begin{align}
\beta_p
=\frac{E[(M_XZ)(M_XY)]}{E[(M_XZ)(M_XW)]}
=\tau+\frac{E[(M_XZ)(M_X\varepsilon)]}{E[(M_XZ)(M_XW)]}. \label{eq:iv_projection_bias}
\end{align}
Conditional validity of $Z$ does not by itself set the last numerator to zero, which can be shown similarly as \eqref{eq:linear_projection_bias}. Thus ordinary 2SLS adjustment need not recover $\tau$; its probability limit is the restricted estimand $\beta_p$.

Nevertheless, $\tau$ can be identified through a control function construction.\footnote{\label{footnote:identification1} This is not the only route. In this linear triangular model, $\tau$ is also identified by $E[\partial_Z E(Y\mid X,Z)]/E[\partial_Z E(W\mid X,Z)]=\tau$.} If $Z$ is independent of $(\varepsilon,\eta)$ conditional on $X$, then $W$ is independent of $\varepsilon$ conditional on $(X,\eta)$. Taking $\widetilde X=(X,\eta)$ and imposing sufficient local variation in $W$ conditional on $(X,\eta)$, together with the required smoothness, the flexible control function estimand is $\beta_{np}=E[\partial_W E(Y\mid W,X,\eta)]=\tau$. Theorem \ref{thm2} later formalizes the construction when the first stage is nonseparable and $\eta$ is unobserved. 

\begin{example}
    Consider a linear triangular model relating the individual wage to school attendance. In an influential paper that studies the causal impact of compulsory school attendance on earnings, \cite{angrist1991does} use quarter of birth (\textit{qbirth}) as an instrument for educational attainment (\textit{totaledu}) in wage equations, based on the observation that school entry requirement and the compulsory schooling laws compel students born in the end of the year to attend school longer than students born in other months. However, this approach is subject to critique that \textit{qbirth} may not be truly exogenous because, for example, family socioeconomic circumstances could be related to conception
timing, which in turn affects quarter of birth. Therefore, researchers may consider including the parents' income (\textit{parinc}) as a control. The heuristic model can be specified as follows:
\begin{align*}
    log(wage) &= \alpha + \tau \cdot totaledu + \xi \cdot parinc + \varepsilon, \\
    totaledu &= \pi_0 + \pi_1 \cdot qbirth +\pi_2 \cdot parinc +
    \eta.
\end{align*}
Parents' income may also be related to unobserved socioeconomic determinants of wages, making it a potentially endogenous control. In that illustrative scenario, linear IV adjustment need not recover $\tau$, whereas the flexible control function estimand can do so under the maintained assumptions. 
\end{example} 

\subsection{Nonlinear binary outcome model}
The issue is not specific to linear projection. Consider the threshold crossing model
\begin{align}
Y=1\{\tau W+X'\xi+\varepsilon>0\}. \label{eq:probit_preview}
\end{align}
The structural average response is the average derivative of the response probability induced by the threshold model. A usual probit QMLE imposes a restricted index and disturbance distribution and therefore targets a restricted estimand $\beta_p$. When the conditional distribution of $\varepsilon$ varies with $X$, that estimand need not equal the structural response. The flexible probability derivative estimand
\begin{align}
\beta_{np}=E[\partial_W P(Y=1\mid W,X)] \label{eq:binary_flexible}
\end{align}
recovers the structural average response under the conditions developed in Section \ref{sec:nonseparable}. 


The examples distinguish three objects. The structural target is $\beta_0$. A restricted parametric procedure targets $\beta_p$, while a flexible conditional mean procedure targets $\beta_{np}$. Under the paper's conditional independence, conditional treatment variation, and smoothness conditions, we have $\beta_{np}=\beta_0.$
By contrast, the further equality $\beta_p=\beta_0$ is not guaranteed under the same set of conditions and requires the restrictions imposed by the parametric method.

\section{Main results}
\label{sec:nonseparable}
\subsection{Structural average response}

To investigate the impact of endogenous controls in a general setting, we consider a broad class of parametric models as \eqref{eq:parametric}. Let $m(W,X,\varepsilon)$ be a nesting nonseparable and nonparametric model, 
\begin{align}
    Y = m(W,X,\varepsilon). \label{eq:nesting} 
\end{align}

A similar nonparametric and nonseparable model has been studied by \cite{altonji2005cross}, except that the external variable they condition on do not enter the outcome equation.\footnote{Our motivation differs because empirical researchers often seek to include $X$ as controls in the outcome equation.} For continuous $W$ and $Y$, define the structural average response
\begin{align}
    \beta_0 = \int \beta_0(w,x)d F(w,x), \quad
    \beta_0(w,x) =\int{\partial_w m(w,x,\epsilon)} f_{\varepsilon\mid W=w,X=x}(\epsilon)d\epsilon. \label{eq:general_response}
\end{align}
This averages the structural response over the observed joint distribution of $(W,X)$.

If $Y$ is a binary outcome, we can write $
    m(W,X,\varepsilon) =1\{m^*(W,X,\varepsilon)>0\}
$, 
where $m^*$ is implicitly defined. Following \cite{altonji2005cross}, partition $\varepsilon=(u,v)$ and define $u^*(W,X,v)$ as the unique threshold solving $m^*(W,X,u^*,v)=0$. Under the monotonicity and differentiability conditions stated in Theorem \ref{thm1}, the structural average response for a binary outcome is defined as
\begin{align}
    \beta_0^* = &\int \beta_0^*(w,x)dF(w,x), \label{eq:binary_response} \\
     \beta_0^*(w,x) = &\int -\partial_wu^*(w,x,v) f_{u,v\mid W=w,X=x}(u^*(w,x,v),v)dv. \nonumber
\end{align}

For a binary treatment, we can define a corresponding set of parameters by replacing the derivatives above with the differences. Identification proceeds analogously in the binary treatment case, as illustrated by the binary treatment example in Section~\ref{sec:motivation}. To highlight the main idea and simplify the exposition, we restrict attention to the continuous treatment case in the main results below.

\begin{example}
    \label{eg3}
        Consider a parametric nonlinear and nonseparable model $Y = h(W,X,\varepsilon;\theta) = 1\{ \tau W + X'\xi  +  \varepsilon > 0 \} $ where $\theta = (\tau,\xi)$ and $\varepsilon|W,X\sim N(\mu(W,X),\sigma(W,X)^2)$; $\mu(W,X)$  and  $\sigma(W,X)^2$ are the conditional mean and variance since we don't restrict the dependence between $\varepsilon$ and $(W,X)$. Suppose we can partition $\varepsilon$ as $ \varepsilon= u+v $, and $v$  is also Gaussian.   Note that $-u^*(W,X,v) =  \tau W + X'\xi  +v $, then $-\partial_wu^*(w,x,v)=\tau$ and 
        \begin{align}
            \beta_0^*(w,x) =& \tau\int f_{u,v|W=w,X=x}(-(\tau w+x'\xi+v),v)dv \nonumber \\ 
            =& \tau\int f_{\varepsilon,v|W=w,X=x}(-(\tau w+x'\xi),v)dv \nonumber \\
            =&  \tau f_{\varepsilon|W=w,X=x}(-(\tau w+x'\xi)) = \frac{\tau}{\sigma(w,x)} \phi\left( \frac{\tau w+x'\xi -\mu(w,x)}{\sigma(w,x)}\right) \label{eq:probit}
        \end{align}
        where the second equality uses $\varepsilon=u+v$ and the change of variables in the joint density. If $\varepsilon\mid W,X\sim N(0,1)$, then $\beta_0^*(w,x)=\tau\phi(\tau w+x'\xi)$, and $\beta_0^*=\tau E[\phi(\tau W+X'\xi)]$ reduces to the usual probit average partial effect.
\end{example}

\subsection{Identification under conditional independence} \label{sec:case1}
Let $\widetilde{X}$ denote a collection of control variables such that $X\subseteq \widetilde{X}$. We consider the scenario where $\varepsilon$ is conditionally independent of $W$ given $\widetilde{X}$, but the dependence between $X$ and $\varepsilon$ is unrestricted. 

\begin{assumption} \label{assum:cond_ind}
$\varepsilon \indep W \ |\  \widetilde{X}$.
\end{assumption}

Assumption \ref{assum:cond_ind} removes variation in the conditional distribution of $\varepsilon$ with $W$ after $\widetilde X$ is fixed. It does not require the controls in $\widetilde X$ to be exogenous. It does not, however, guarantee that $W$ varies after those controls are fixed. If $W$ is determined by $\widetilde X$, the conditional independence may hold vacuously and the response remains unidentified because no conditional treatment comparison is available.

The second identifying ingredient is conditional treatment variation. For a continuous treatment, the response is averaged at treatment values lying in the interior of the conditional support. For a binary treatment, the corresponding requirement is conditional overlap. Let
\[
\mathcal W(\widetilde x)
=
\operatorname{supp}(W\mid\widetilde X=\widetilde x).
\]

\begin{assumption}[Conditional treatment variation] \label{assum:local_support}
For almost every $\widetilde x$,
    \[
    P\!\left(
    W\in\operatorname{int}\mathcal W(\widetilde x)
    \mid \widetilde X=\widetilde x
    \right)=1.
    \]
\end{assumption}

Assumption~\ref{assum:local_support} is a local support condition: almost every observed treatment value can be perturbed in a neighborhood while $\widetilde X$ is held fixed. For binary treatment, an analogue is a conditional support conditions such as $0<P(W=1\mid\widetilde X=\widetilde x)<1$, where both treatment states can be compared at the same value of $\widetilde X$. In this sense, Assumption \ref{assum:local_support} is the relevant conditional rank or variation condition. It is weaker than rectangular support, which would require a common treatment range across control values. Smoothness is additionally imposed below when the target involves the derivative.

\begin{remark}
    A related rank condition in the literature for identification in nonseparable models is the \emph{measurable separability}; see \cite{florens2008identification}. The variables $W$ and $\widetilde X$ are measurably separated if $a(W)=b(\widetilde X)$ almost surely for measurable functions $a$ and $b$ implies that their common value is almost surely constant. This condition rules out nonconstant common measurable components of treatment and controls. Like Assumption~\ref{assum:local_support}, it addresses whether treatment and control variation can be distinguished for identification. Measurable separability has been used together with additional model restrictions to establish identification of treatment effects. Our analysis instead imposes local conditional support, which directly
    provide the comparisons needed to identify conditional derivatives.
\end{remark}

\begin{theorem}
    \label{thm1} Suppose Assumptions \ref{assum:cond_ind} and \ref{assum:local_support} hold. 
    \begin{enumerate}
        \item If $Y$ is continuous, suppose $m(w,x,e)$ is differentiable in $w$ on a neighborhood of the relevant treatment values and its derivative is dominated by an integrable function on that neighborhood. Then, $\beta_0 = E\left[\partial_WE\left(Y\mid W,\widetilde{X}\right)\right]$.
        \item If $Y$ is binary, suppose $m^*(w,x,u,v)$ is strictly increasing in $u$ and has a root $u^*$; $m^*(w,x,u,v)$ is continuously differentiable in $w$ and $u$; $\partial_u m^*(w,x,u^*,v)\neq 0$; and $f_{u,v|\widetilde X}$ is continuous in $u$ at $u^*(w,x,v)$. Suppose  $\left| \partial_w u^*(w,x,v)\, f_{u,v\mid \widetilde X=\widetilde x} \bigl(u^*(w,x,v),v\bigr) \right|$ is dominated by an integrable envelope. Then, $\beta_0^* = E\left[\partial_W E\left(Y\mid W,\widetilde X\right)\right]$.
    \end{enumerate}
\end{theorem}

Theorem \ref{thm1} is the main result behind the motivating examples. Identification by conditional derivatives under conditional independence is familiar; its role here is to clarify when dependence between controls and structural disturbances does and does not invalidate adjustment. Conditional independence permits the structural disturbance to depend on $X$ but holds its conditional distribution fixed as $W$ varies. Local conditional support supplies the treatment variation needed to identify the derivative from observables, while smoothness and domination justify differentiation under the integral. A proof is in the Appendix.

For a binary treatment, differentiation is replaced by a difference and local support of Assumption~\ref{assum:local_support} is replaced by a conditional overlap condition. The same conditioning argument identifies the structural response from $E(Y\mid W=1,\widetilde X)-E(Y\mid W=0,\widetilde X)$, averaged over the relevant distribution of the controls. No smoothness in $W$ is needed in that case.

The result points out a promising route to address potentially endogenous controls and adds an extra justification for employing a flexible functional form. However, the result does not imply that potentially endogenous controls are harmless in every application. Rather, it shows that flexible adjustment recovers the average response when conditional validity and sufficient conditional treatment variation hold. The weak conditional variation design in Section~\ref{sec:sim} illustrates why this distinction matters.

\begin{remark}
    One may view the result as conditioning on $X$. With continuous and multivariate controls, however, literal comparisons at each exact value of $X$ are infeasible and can obscure the underlying estimation problem. Our approach instead differentiates the conditional mean over the observed support, replacing infeasible cell-by-cell comparisons with the substantially weaker requirement of sufficient conditional treatment variation.
\end{remark}

\noindent \textbf{Example 3 continued.} 
        For the binary model $Y = 1\{ \tau W + X'\xi  + \varepsilon>0 \} $, suppose $\varepsilon|W,X \sim N(\mu(X),\sigma(X)^2)$, i.e. Assumption~\ref{assum:cond_ind} holds for this model with $\widetilde{X} = X$. Following the calculation as \eqref{eq:probit}, the true average response is given by
        \begin{align*}
           \beta_0^*= E[\beta_0^*(W,X)] =E\left[\frac{\tau}{\sigma(X)} \phi\left( \frac{\tau W+X'\xi -\mu(X)}{\sigma(X)}\right)\right].
        \end{align*}
        By Theorem \ref{thm1}, $\beta_0^*$ is identified as $E[\partial_W P(Y=1\mid W,X)]$ when local conditional support and the required smoothness conditions hold. Indeed, this derivative equals $E\left[\frac{\tau}{\sigma(X)}\phi\left(\frac{\tau W+X'\xi-\mu(X)}{\sigma(X)}\right)\right]$ under Assumption~\ref{assum:cond_ind}. Standard probit instead imposes $P(Y=1\mid W,X)=\Phi(\tau W+X'\xi)$; its QMLE therefore need not recover the structural response when the conditional disturbance distribution varies with $X$.

\subsection{Identification with IV}
\label{sec:case2}

When Assumption~\ref{assum:cond_ind} is not plausible given observed $X$, an excludable instrument may be needed for identification. Given such an instrument and its conditional validity, we construct a control variable $V$ such that Assumption~\ref{assum:cond_ind} holds conditional on $\widetilde X=(X,V)$. This control function construction follows \cite{imbens2009identification}. Here, however, its role differs in two respects. Because instrument validity is conditional on $X$ and $X$ may itself be endogenous, conditional independence is recovered only after conditioning jointly on $(X,V)$. In addition, identification requires sufficient conditional treatment variation given $(X,V)$.

The researcher starts from \eqref{eq:parametric}, nested in the nonparametric and nonseparable \eqref{eq:nesting}, with structural target $\beta_0$. Suppose an exogenous and excludable variable $Z$ generates the reduced form
\begin{align*}
    W=q(Z,X,\eta)
\end{align*}
where $\text{dim}(\eta) = \text{dim}(W) = 1$.\footnote{As discussed in \cite{imbens2007nonadditive}, the same dimension of the treatment and the endogenous error term is essential for recovering all endogeneity in $\eta$ through the inverting procedure.} The exogeneity of the IV is represented by the conditional independence in the following assumption.

\begin{assumption} \label{assum:cond_ind_iv} $\ Z \indep  (\varepsilon,\eta) \ | \  X.$
\end{assumption}
Note that we don't impose the exogeneity condition of $X$ with respect to either $\varepsilon$ or $\eta$, yet $X$ is allowed in both outcome and the reduced form equations in a nonseparable way. The requirement for IV is relaxed, and $Z$ may well be dependent on $X$, which motivates the inclusion of $X$ in the model.

From a control function perspective, the goal is to find $V$ such that 
\begin{equation}
    \varepsilon \indep W \ | \ \widetilde{X} = \{X,V\},
    \label{eq:controlv}
\end{equation}
and $W$ has sufficient conditional variation given $\widetilde X$ as required by Assumption \ref{assum:local_support}. Because the IV restriction is conditional on $X$, the relevant conditional rank transform is $F_{\eta\mid X}(\eta)$ rather than the unconditional $F_\eta(\eta)$. 

\begin{assumption} \label{assum:v_eta}
    (i) $q(Z,X,\eta)$ is strictly increasing in $\eta$ almost surely; (ii) The conditional CDF $F_{\eta|X}(a)$ is continuous and strictly increasing in $a\in supp(\eta)$ almost surely. 
\end{assumption}

Assumption \ref{assum:v_eta}(i) fixes the direction of monotonicity so that $F_{W\mid Z,X}(W)=F_{\eta\mid X}(\eta)$. If $q$ is strictly decreasing, the corresponding control is the reversed transform $1-F_{W\mid Z,X}(W)$. Assumption \ref{assum:v_eta}(ii) ensures that the recovered rank captures the information in $\eta$ conditional on $X$. 

\begin{theorem}    \label{thm2}
    Suppose Assumption \ref{assum:cond_ind_iv} holds for the nonseparable model \eqref{eq:nesting}. Then, 
    \begin{itemize}
        \item[(i)] $W$ is independent of $\varepsilon$ conditional on $(\eta,X)$.

        \item[(ii)] If, additionally, Assumption \ref{assum:v_eta} holds, then $F_{W|Z,X}(W)= F_{\eta|X}(\eta)$, and condition \eqref{eq:controlv} is satisfied with $V = F_{W|Z,X}(W)$.
    \end{itemize}
\end{theorem}

Theorem \ref{thm2} therefore provides the control needed to establish Assumption~\ref{assum:cond_ind} with $\widetilde X = (X,V)$. Applying Theorem \ref{thm1} additionally requires local support of $W$ given $(X,V)$ and the stated smoothness conditions. The proof is in the Appendix.\footnote{Although $V$ does not enter $m(W,X,\varepsilon)$, define $\widetilde m(W,X,V,\varepsilon)=m(W,X,\varepsilon)$ when applying Theorem \ref{thm1}.}

\section{ A specification test for endogenous controls}
\label{sec:test}
This section formalizes a specification test for endogenous controls to guide the choice between restrictive and flexible functional forms. We emphasize that the test differs from a conventional specification test: even when the structural model in \eqref{eq:parametric} is correctly specified, the restricted parametric estimand need not coincide with the structural parameter of interest because we do not restrict the dependence between the controls and the disturbance term. When this discrepancy arises, the restricted parametric and flexible estimands differ, motivating the test developed in this section. If \eqref{eq:parametric} is itself misspecified, which is outside the scope of this paper but common in practice, the proposed test may also reject for conventional functional-form misspecification. Thus, in that case, rejection indicates a consequential discrepancy between the restricted and flexible specifications, but does not by itself distinguish whether the discrepancy arises from endogenous controls or from misspecification of the structural model.

Throughout the section, we restrict our attention to the case under correct specification of \eqref{eq:parametric}, and we focus on observable $\widetilde X=X$ and an i.i.d. sample of size $n$.\footnote{The analysis can be extended to Section \ref{sec:case2}, where $V$ is estimated, at the cost of additional first-step terms.} We consider the null and the alternative hypotheses as follows:
\begin{align*}
        H_0:\ X \text{\ is \ exogenous}, \quad H_1: \ X \text{\ is \ endogenous}.
\end{align*}
The null and alternative hypotheses are specified loosely to accommodate a broad class of models. The meaning of the null and the alternative varies over the specification of $  h(W,X,\varepsilon;\theta)$. For example, in the linear model, $E[X\varepsilon] = 0$ is necessary and sufficient for the null while $X\indep \varepsilon$ is not necessary. However, in the endogenous probit model of Example \ref{eg3}, $X\indep \varepsilon$ is necessary and sufficient for the null while $E[X\varepsilon] = 0$ is not sufficient. We will make the hypotheses concrete in a moment when we introduce the assumptions for the estimators under consideration.

To represent both binary and continuous outcomes, we denote the average response as $\beta_0:=E[\partial_Wh(W,X,\varepsilon;\theta)]$ for both cases throughout this section. Let $\hat{\beta}_p$ denote the parametric estimator under model \eqref{eq:parametric}, but treating $X$ as exogenous, i.e. imposing $H_0$. For the proposed nonparametric method, we consider an average derivative estimator $\hat{\beta}_{np}$: let $g(W,X) := E[Y|W,X]$, and
\begin{align}
    \hat{\beta}_{np} = \frac{1}{n}\sum_{i=1}^n\partial_W\hat{g}(W_i,X_i) \label{eq:ave_deri_est}
\end{align}
where $\hat{g}$ is a nonparametric estimator for $g$. 

Let $\beta_p$ and $\beta_{np}$  denote the estimands of the restricted parametric method and the flexible method, respectively. With an appropriate choice of $\hat{g}$ and sufficient regularity conditions, the average derivative estimator is consistent for $\beta_{np} = E[\partial_Wg(W,X)]$ (e.g., see Example 3 of \citealp{newey1994asymptotic}). Under the conditions of Theorem 1, $\beta_{np} = \beta_0$, regardless of $H_0$ or $H_1$. Meanwhile, under the same conditions, the limit $\beta_p$ may change across data generating processes because the parametric model is correctly specified up to the exogeneity restrictions on the controls. 

If, under $H_0$, $\beta_{p} = \beta_0$, and both estimators, $\hat{\beta}_p$ and $\hat{\beta}_{np}$, are asymptotically linear with some influence functions $\varphi$ and $\phi$ respectively, then we can define $\psi = \varphi - \phi$, and write 
\begin{align*}
    \sqrt{n}\left(\hat{\beta}_p-\hat{\beta}_{np}\right) =\frac{1}{\sqrt{n}}\sum_{i=1}^n \psi(W_i,X_i)+o_P(1),
\end{align*}
which, under common regularity conditions, is asymptotically normal with the asymptotic variance $ V = \lim_{n\to\infty} Var\left(\frac{1}{\sqrt{n}}\sum_{i=1}^n \psi(W_i,X_i)\right)$. When the parametric estimator $\hat\beta_p$ is efficient under the null, we can appeal to the classic result of \cite{hausman1978specification} to express the $V$ as the difference of the asymptotic variances of $\hat{\beta}_{np}$ and $\hat{\beta}_p$. In general, however, assuming efficiency of the parametric estimator is not desirable. Moreover, since $\varphi(W,X)$ depends on specific parametric estimator, and $\phi(W,X)$ may depend on extra infinite dimensional nuisance parameter estimation, the analytical expression for $V$ is not straightforward. Alternatively, we can resort to a bootstrap approach, which is a common and general approach for inference that involves semiparametric estimation; see, for example, \cite{chen2003estimation}.

Therefore, with $\theta_0 =  \beta_p - \beta_{np}$ and $\hat\theta = \hat{\beta}_{p} - \hat\beta_{np}$, we consider bootstrapping the non-studentized statistic $ \sqrt{n}\left(\hat{\theta}-{\theta}_0\right)$. Let $J_n^*$ denote the empirical bootstrap CDF of $\sqrt{n}(\hat\theta^*-\hat \theta)$ where $\hat\theta^*$ is the bootstrap version of $\hat\theta$. Define the bootstrap confidence region
$$ \mathcal{B}_n(\alpha) = \{\theta\in \Theta: \hat\theta  - n^{-1/2} \inf \{t: J_n^*(t)\geq 1- \alpha/2 \}\leq \theta \leq   \hat\theta  - n^{-1/2}\inf \{t: J_n^*(t)\geq \alpha/2 \}\}.$$ 
A brief algorithm for implementation of the test is given as follows.

\begin{algorithm}
    \begin{itemize}
        \item Step 1: Based on the parametric model \eqref{eq:parametric}, obtain $\hat\beta_p$ treating $X$ as exogenous. Obtain $\hat{\beta}_{np}$ as \eqref{eq:ave_deri_est} with some nonparametric estimator $\hat{g}$ for $E[Y|W,X]$. Let $\hat\theta = \hat{\beta}_{p} - \hat\beta_{np}$.
    
        \item Step 2: Resample independently with replacement from $P_n$ with equal probability $1/n$ for each observation. Obtain the bootstrap sample $\{Y_i^*,W_i^*,X_i^*\}_{i=1}^n$. 

        \item Step 3: Obtain $\hat{\beta}_{p}^*$, $\hat g^*$, $\hat\beta_{np}^* = \frac{1}{n}\sum_{i=1}^n \partial_W\hat g^*(W_i^*,X_i^*)$, and $\hat\theta^* = \hat{\beta}_{p}^* - \hat\beta_{np}^*$ using $\{Y_i^*,W_i^*,X_i^*\}_{i=1}^n$.

        \item Step 4: Repeat Steps 2 - 3 for a sufficiently large $B$ times and denote each bootstrap realization using subscript $b$. Obtain the empirical bootstrap CDF $J^*(t) = \frac{1}{B}\sum_{b=1}^B 1\{\sqrt{n}(\hat{\theta}^*_b - \hat\theta)\leq t\}$. Obtain $\mathcal{B}_n(\alpha)$. Reject the null if $0\notin \mathcal{B}_n(\alpha)$.
    \end{itemize}
\end{algorithm}

We introduce a set of high level conditions in terms of $\hat{\beta}_p$, $\hat{\beta}_{np}$, as well as their limits. Let $P^*$ denote the bootstrap probability measure conditional on the sample $\{Y_i,W_i,X_i\}_{i=1}^n$.

\begin{assumption}
    \label{assum:test} (i) Under $H_0$, $\beta_p = \beta_0$; Under $H_1$, $\beta_p = \beta_0 + B_n$ for some real sequence $B_n$ with $\sqrt{n}|B_n|\to\infty$. (ii) There exists $\varphi (W,X)$ with $E[\varphi (W,X)] = 0$, 
    \begin{align}
        \hat{\beta}_p - \beta_p= & \frac{1}{n}\sum_{i=1}^n\varphi (W_i,X_i) + o_P(n^{-1/2}). \label{eq:asym_linear_p} \\
       \hat{\beta}_p^* - \hat{\beta}_p 
        =& \frac{1}{n}\sum_{i=1}^n \varphi(W_i^*,X_i^*) - \varphi(W_i,X_i) + o_{P^*}(n^{-1/2}). \label{eq:asym_linear_p_boot}
    \end{align}
    (iii) There exists $\phi (W,X)$ with $E[\phi (W,X)] = 0$,
    \begin{align}
         \hat{\beta}_{np} - \beta_{np}
        =& \frac{1}{n}\sum_{i=1}^n \phi (W_i,X_i)  + o_P(n^{-1/2}). \label{eq:asym_linear_np} \\
        \hat{\beta}_{np}^* - \hat{\beta}_{np}
        =& \frac{1}{n}\sum_{i=1}^n \phi(W_i^*,X_i^*)- \phi(W_i,X_i) + o_{P^*}(n^{-1/2}). \label{eq:asym_linear_np_boot}
    \end{align}
    (iv) For $\psi = \varphi-\phi$, $0<\text{Var}(\psi(W,X))<\infty$.
\end{assumption}

Assumption \ref{assum:test} (i) operationalizes the null and alternative hypotheses. For example, under the linear model \eqref{eq:linear}, $\beta_0 = \tau$ and $\beta_p$ is the first element of the linear projection parameter $\gamma_{LP}$, thus $B_n$ is the first element of $E[DD']^{-1}E[DE[\varepsilon|X]]$ where $D = (W,X')'$. The upper bound in Assumption \ref{assum:test}(iv) is a common finite variance condition for applying the central limit theorem. The lower bound excludes the case where the parametric estimator and the average derivative estimator agrees. 

Assumption \ref{assum:test} (ii) and (iii) impose asymptotic linear representations for the estimators and their bootstrap counterparts. For parametric estimators, asymptotic linear representation is very common. For example, it holds for M-estimators with twice continuously differentiable objective functions. For some estimators with non-smooth objective functions, e.g. quantile estimators, there are also existing results established under extra regularity conditions. The asymptotic linear representation for the average derivative estimator follows from the general results for semiparametric estimators. \cite{newey1994asymptotic} provides a general approach for deriving the influence function for semiparametric estimators, and Example 3 of the same paper exemplifies the average derivative estimator using series approximation for the conditional expectation. See the next section for a concrete implementation of this estimator in simulation. 

The asymptotic linear representations for bootstrap estimators expect slightly stronger regularity conditions. To see how this representation can be justified, we consider the process of the recentered scores $S_n(g):= \frac{1}{\sqrt{n}}\sum_{i=1}^n \left( \partial_W g(W_i,X_i) - E[ \partial_Wg(W_i,X_i)] \right)$ and the bootstrap process $S_n^*(g) = \frac{1}{\sqrt{n}}\sum_{i=1}^n \left( \partial_W g(W_i^*,X_i^*) - \partial_Wg(W_i,X_i)\right)$. Also, define the pathwise derivative of $S_n(g)$ in the direction of $\tilde g - g$, evaluated at $g$, as $\Gamma(g)[\tilde g - g]$. In a semiparametric GMM framework, Theorem B of \cite{chen2003estimation} shows that under a slightly strengthen set of sufficient conditions for asymptotic normality of the GMM estimator, \eqref{eq:asym_linear_np_boot} can be obtained by 
\begin{align}
    \Vert\hat g^* - \hat g\Vert_{\infty} = &o_{P^*}(n^{-1/4}), \label{eq:nprate}\\
    S_n^*(\hat g) + \Gamma(\hat g)[\hat g^* - \hat g] =& \sum_{i=1}^n \left(\phi(W_i^*,X_i^*) - \phi(W_i,X_i) \right)+ o_{P^*}(1). \label{eq:equi_and_pathwise}
\end{align}
Condition \eqref{eq:nprate} is also assumed in \cite{chen2003estimation} and holds for, for example, series estimators(\citealp{NEWEY1997147,chen2015optimal}) and kernel regression estimators(\citealp{hall1991convergence}). For Condition \eqref{eq:equi_and_pathwise}, we further define, for a fixed $g$,
\[
M_n(g) := \tfrac{1}{\sqrt n}\sum_{i=1}^n \partial_W g(W_i,X_i), \qquad
M_n^*(g) := \tfrac{1}{\sqrt n}\sum_{i=1}^n \partial_W g(W_i^*,X_i^*),
\]
so that $S_n(g) = M_n(g) - \sqrt n\,E[\partial_W g(W,X)]$ and
$S_n^*(g) = M_n^*(g) - M_n(g)$. Since
$\sqrt n(\hat\beta_{np}^* - \hat\beta_{np}) = M_n^*(\hat g^*) - M_n(\hat g)$,
adding and subtracting $M_n^*(\hat g)$ and $\Gamma(\hat g)[\hat g^*-\hat g]$ yields
\begin{align*}
\sqrt n\left(\hat\beta_{np}^* - \hat\beta_{np}\right)
= \underbrace{S_n^*(\hat g) + \Gamma(\hat g)[\hat g^* - \hat g]}_{=:A_n^*}
+ \underbrace{M_n^*(\hat g^*) - M_n^*(\hat g) - \Gamma(\hat g)[\hat g^* - \hat g]}_{=:R_n^*}.
\end{align*}
The term $R_n^*$ is the bootstrap analogue of the remainder in the pathwise
expansion of $g \mapsto M_n^*(g)$ around $\hat g$. It is $o_{P^*}(1)$ under
stochastic equicontinuity of $M_n^*(\cdot)$ over shrinking neighborhoods of
$\hat g$ together with the rate condition \eqref{eq:nprate}, which makes the
second order term in the expansion $O_{P^*}(\sqrt n \,\|\hat g^*-\hat g\|_\infty^2)
= o_{P^*}(1)$. For $A_n^*$, under regularity conditions on $\Gamma(\hat g)[\hat g^* - \hat g]$
under $P^*$, Lemma 5.1 of \cite{newey1994asymptotic} delivers
\eqref{eq:equi_and_pathwise}, with $\phi$ the influence function of the average
derivative estimator given in Example 3 of the same paper.

The following result shows that this bootstrap confidence region is asymptotically valid.

\begin{theorem}
\label{thm:3}
Under Assumptions \ref{assum:test} and conditions of Theorem \ref{thm1}, as $n\to\infty$, $ P\{0 \in \mathcal{B}_n(\alpha) | H_0\} \to  1-\alpha$ and $ P\{ 0 \notin \mathcal{B}_n(\alpha)|H_1 \}  \to 1$.
\end{theorem}

Theorem \ref{thm:3} formalizes the Hausman-type principle for the endogenous control test. Unlike a conventional specification test, the rejection here can occur even when the parametric model is correctly specified. A proof can be found in the Appendix.

\section{ Simulation}
\label{sec:sim}

The simulations are conducted around four questions: How large is the bias from omitting the control $X$? Does restricted inclusion eliminate it? Does the flexible ADE recover $\beta_0$, and at what variance cost? What happens when conditional treatment variation becomes weak? The first three questions are addressed with DGP (1) and DGP (2), while DGP (3) isolates the role of conditional treatment variation.


For concreteness, we consider an implementation using series estimator to estimate conditional mean functions and their derivatives, based on our constructive identification results. Suppose $(W,\widetilde{X})$ is $d$-dimensional. For some given $K$, we define a power series $\{\tilde{g}_k\}_{1\le k\le K}$ to approximate $g(W,\widetilde{X})=E[Y|W,\widetilde{X}]$ by
$    \tilde g := \sum_{k=1}^K\tilde{g}_k'\pi_k = G' \Pi$,
where $G = (\tilde{g}_1',...,\tilde{g}'_K)'$; for each $k$, 
$\tilde{g}_k= \text{vec}\left\{\prod_{j=1}^d q_j^{\lambda_j}: \sum_{j=1}^d\lambda_j = k, (q_1,...,q_d) = \left(l_1(W),l_2(\widetilde{X}_1) ,...,l_d(\widetilde{X}_{d-1})\right)\right\}$, and $l=(l_1,...,l_d)$ is some transformation function differentiable to all orders with bounded derivatives and has $\text{det}(\partial_z l(z))$ bounded away from zero. $\Pi = (\pi_1',...,\pi_K)'$ are the corresponding linear projection parameters. The associated average derivative estimator is defined as
$$\hat{\beta}_{np}:= \frac{1}{n}\sum_{i=1}^n \partial_W \left[G(W_i,\widetilde{X}_i)'\hat\Pi\right],$$
where $\hat\Pi$ are the (penalized) least square estimators of $\Pi$. For our implementation in this section, we fix $K=3$ and set $l(x)=x$.

The first DGP covers the scenario where both the treatment and the controls are endogenous, while the treatment is conditionally independent of the unobserved determinants:
\begin{align*}
   \text{DGP(1):\ } Y & = \tau W +  X_1 + X_2+U, \ \ \ W  = \frac{1}{2}\exp(a) + N(0,1) \\
                        X_1 & = a + \frac{1}{2}\exp(p), \ \ \ X_2  = p, \ \ \    U = \frac{1}{2}\exp(b) + \frac{1}{2}\exp(q) + N(0,1),
\end{align*}
where $\tau= 1$; $(a,b)$ and $(p,q)$ are independent of each other, and each are jointly normal with mean zero, variance one, and covariance $\rho$; $N(0,1)$ denotes a random draw from a standard normal distribution, independent of $(a,b)$ and $(p,q)$. We observe that (1) $X = (X_1,X_2)'$ is relevant for both $Y$ and $W$; (2) $W$ and $X$ are dependent on $U$; (3) conditional on $X$, $W$ is independent of $U$; and (4) the additive normal innovation in $W$ gives local treatment variation conditional on $X$. As a result, linear projection parameters do not identify $\tau$. Meanwhile, $\tau$ is also the average response to $W$, and it is nonparametrically identified by Theorem \ref{thm1}. 

The second DGP covers the scenario where the IV is only conditionally valid, and the control is endogenous:
\begin{align*}
\text{DGP(2):\ }  Y & =
 \tau W +  X_1 + X_2 +  U, \ \ \ W = X_1 + X_2 + Z + \eta \\
        X_1 & = a + \frac{1}{2}\exp(p), \ \ \  X_2  = p, \ \ \ Z  = \frac{1}{2}\exp(a) + N(0,1) \\
        \eta &= \frac{1}{2}\exp(\xi) , \ \ \ U = \frac{1}{2}\exp(b) + \frac{1}{2}\exp(q) + \frac{1}{2}\exp(\zeta)  + N(0,1),
\end{align*}
where $(\xi,\zeta)$ are also jointly normal with mean zero, variance one, and covariance $\rho$. We observe that (i) $W$ is not conditionally independent of $U$ given $X$; (ii) $Z$ is a valid IV only conditional on $X$, although $X$ is endogenous; and (iii) the additive normal innovation in $Z$, and hence in $W$, provides local treatment variation conditional on $(X,\eta)$. 
As a result, $\tau$ is not identified by the usual IV projection, while $\tau$ is again, as an average response, nonparametrically identified: constructive identification is given by the control function approach of Theorem~\ref{thm2}. 

\begin{table}[htbp]
\centering
\begin{threeparttable}
\caption{Comparing different estimators under endogenous controls}
\label{tab:simulation_merged}
\begin{tabular}{ccccc ccc}
\toprule
& & \multicolumn{3}{c}{DGP(1)} & \multicolumn{3}{c}{DGP(2)} \\
\cmidrule(lr){3-5} \cmidrule(lr){6-8}
$\rho$ & Monte Carlo 
& OLS & OLS & Proposed 
& IV & IV & Proposed  \\
& Statistics 
& w.o.\ X & w.\ X & Method
& w.o.\ X & w.\ X & Method \\
\midrule
\multirow{3}*{0.75}
& Bias & 0.732 & 0.156 & 0.013 & 0.528 & 0.156 & 0.020 \\
& SD   & 0.091 & 0.080 & 0.048 & 0.057 & 0.083 & 0.063 \\
& MSE  & 0.545 & 0.031 & 0.002 & 0.282 & 0.031 & 0.004 \\
\addlinespace
\multirow{3}*{0.5}
& Bias & 0.588 & 0.088 & 0.004 & 0.424 & 0.089 & 0.001 \\
& SD   & 0.083 & 0.063 & 0.054 & 0.052 & 0.068 & 0.069 \\
& MSE  & 0.352 & 0.012 & 0.003 & 0.182 & 0.012 & 0.005 \\
\addlinespace
\multirow{3}*{0}
& Bias & 0.383 & 0.000 & 0.000 & 0.277 & 0.009 & 0.004 \\
& SD   & 0.065 & 0.045 & 0.058 & 0.044 & 0.052 & 0.073 \\
& MSE  & 0.151 & 0.002 & 0.003 & 0.079 & 0.003 & 0.005 \\
\bottomrule
\end{tabular}

\medskip
\parbox{\linewidth}{\footnotesize\textit{Note:} Simulation results are based on $10{,}000$ Monte Carlo replications with sample size $n=1000$. For DGP(2), the conditional CDF is estimated based on a grid of 10 quantiles between 0.05 and 0.95 using the empirical support of $W$.}
\end{threeparttable}
\end{table}

For DGP (1), $\rho$ governs control--error dependence and $\rho=0$ is the benchmark. Table \ref{tab:simulation_merged} compares OLS without $X$, OLS with $X$, and the third order series ADE with $X$. Omitting $X$ is biased throughout. Restricted inclusion removes the bias at $\rho=0$ but leaves increasing bias as $\rho$ rises. The ADE has substantially less bias in these designs, with the expected variance cost in some cases.

For DGP (2), $Z$ is invalid without $X$, while linear IV adjustment with $X$ can still leave nonlinear control related disturbance components. The general implementation uses $\widetilde X=(X,\widehat V)$, where $\widehat V$ estimates $F_{W\mid Z,X}(W)$ on 10 quantile grid points, and then estimates the outcome conditional mean by a third order series. Again, we find the flexible construction sharply reduces bias.

\subsection{Weak conditional treatment variation}
To demonstrate the necessity of sufficient conditional treatment variation, we propose the following additional design:
\begin{align}
Y &= \tau W + \xi X+\varepsilon, \quad a,\nu,U\overset{iid}{\sim}N(0,1)\nonumber \\
X&= a ,  \quad W = \frac{1}{2}\exp(X) + \sigma\nu, \quad \varepsilon = \rho X^2 + U 
\label{eq:dgp3}
\end{align}
with $\tau=\xi=1$, $\rho=0.5$, and $\sigma\in\{1,0.1,0.01,0.001,0\}$. Assumption~\ref{assum:cond_ind} holds because $U$ is independent of $\widetilde X =(W,X)$, so $\varepsilon\indep W\mid X$. For every $\sigma>0$, $W$ retains local conditional support given $X$, but the conditional distribution becomes increasingly concentrated as $\sigma\to0$. The formal support condition therefore holds for each positive $\sigma$ even as effective treatment variation weakens and estimation becomes ill-conditioned. At $\sigma=0$, the conditional support collapses to a singleton, and $\tau$ is not separately identified.

\begin{table}[htbp]
\centering
\caption{Weak conditional treatment variation (DGP (3))}
\label{tab:dgp3}
\begin{tabular}{ccccc}
\toprule
$\sigma$ & OLS bias & ADE bias & ADE SD & ADE RMSE \\
\midrule
1 & 0.274 & 0.000 & 0.032 & 0.032 \\
0.1 & 0.899 & -0.001 & 0.320 & 0.320 \\
0.01 & 0.920 & 0.046 & 3.143 & 3.144 \\
0.001 & 0.924 & -0.033 & 30.754 & 30.753 \\
0 & 0.924 & -25.214 & 1864.308 & 1864.385 \\
\bottomrule
\end{tabular}
\end{table}

Table~\ref{tab:dgp3} illustrates the importance of distinguishing conditional validity from conditional treatment variation. When $\sigma=1$, the ADE is essentially unbiased and precisely estimated. As $\sigma$ decreases, its bias remains small for every positive $\sigma$, consistent with identification under the local-support condition, but its sampling variability increases dramatically: the standard deviation rises from $0.032$ at $\sigma=1$ to $0.320$, $3.143$, and $30.754$ as $\sigma$ falls to $0.1$, $0.01$, and $0.001$, respectively. Thus, although the formal support condition continues to hold for any $\sigma>0$, increasingly concentrated conditional treatment variation makes estimation progressively ill-conditioned. At $\sigma=0$, the conditional support of $W$ collapses and the ADE becomes extremely unstable, reflecting the failure of identification. By contrast, the bias of the restricted OLS estimator persists regardless of the amount of conditional treatment variation governed by $\sigma$. Overall, the results show that conditional independence alone is not sufficient for practically useful estimation: flexible adjustment can eliminate endogenous control bias only when there is also sufficient treatment variation after conditioning.

\section{Application: The China Syndrome Revisit}
\label{sec:example}
In this section, we revisit an empirical study of the import competition effect on the US labor market. During the late 20th and early 21st century, the world has witnessed a drastic surge of imports from the Chinese market. Meanwhile, there was a downturn of import competing manufacturing in certain regions of the US, and it came with a higher unemployment rate and wage inequality in these regions. Therefore, a natural question arises: was the disruption of the local labor market mainly caused by the import competition, or was it rather a result of an overall economic transition in the US? 

In a classic paper by \cite{autor2013china}, the authors answer this question by exploiting the import competition exposure variation across regional markets in the US. The idea is that regions with more initial specialization in labor-intensive industries are more exposed to the Chinese import competition. They measure the \textit{share} of those industries at the start of the observation period and multiply it by the growth in US imports from China (\textit{shift}), which produces a measure of import competition that varies across regions. By taking commuting zones as analysis units, they are able to estimate effects on various labor market outcomes with a reasonable sample size. 

There are two potential sources of endogeneity in this setup. Firstly, the import growth from China could be driven by unobserved local market transitions, such as industrial reallocation that causes a supply shortage. In that case, the growth in US imports from China could be driven by unobserved shocks that also move local outcomes. Secondly, the start-of-period measure of share in labor-intensive industries may also be related to other regional characteristics, such as educational attainment, which also determines the local labor market outcomes. To deal with the first type of endogeneity, they employ a measure of import growth in other high income countries as an exogenous \textit{shift} and multiply it by the same \textit{share} variable to obtain the IV. They further assume that the growth in imports from China is not driven by demand shocks that are shared by the US and other high income countries, so as to ensure the validity of the IV. For the second type of endogeneity, they take two approaches: (1) They first-difference both the treatment and the outcomes so that they are in terms of per capita changes. In a way, this is similar to removing the fixed effects in a linear panel model. (2) They add other start-of-period measures of regional market conditions and demographics to make the exogeneity of the \textit{share} variables more plausible. 

The results in Table 3 of \cite{autor2013china} show that adding more controls reduces the size of the estimated (negative) impact due to Chinese imports. This can be regarded as correcting the bias due to the second type of endogeneity. However, adding more controls may not fully capture the unobserved determinants that jeopardize the exogeneity of the \textit{share} variables, and doing so may introduce extra endogeneity from the control variables. To make the exogeneity condition of the IV more likely to hold while avoiding the extra bias due to endogenous controls, we propose to implement the nonparametric control function approach in Section \ref{sec:case2}. We also illustrate how the endogenous control test works and its practical implications. 

Our approach is based on the exogeneity of the IV \textit{shift} variables\footnote{Suppose the growth in imports from China is partially driven by positive demand shocks too, then the estimates of import-competition impact would look smaller because the positive demand would also generate positive outcomes in the US markets.}. We do not tend to evaluate or improve upon the choice of IV in the original study; instead, we intend to make this approach more robust by allowing more controls to achieve conditional validity of IV while relaxing the exogeneity requirement of the control variables. 

To focus on our main concern, readers are referred to \cite{autor2013china} for a detailed construction of the data. The main outcome of interest, $\Delta L$, is the regional decade change (\% pts) in the share of manufacturing employment; the explanatory variable of focus, $\Delta I\_{\rm US}$, is a measure of changes in Chinese import exposure per worker in each region; and the IV, $I\_{\rm O}$, is the same measure of changes in exposure but with Chinese imports to the US replaced by those to other high income countries. Two periods of decade long first-difference cross sectional data are stacked together as a panel. Five sets of time invariant control variables, $X$, are augmented. The baseline model is given as follows:
\begin{align*}
    \Delta L_{it} = \gamma_t + \tau \Delta I\_{\rm US}_{it} + X_{i}^\prime \xi + e_{it}
\end{align*}
where $\gamma_t$ is a time fixed effect and $e_{it}$ is the stochastic error term. Under the IV conditions and regularity conditions, $\tau$ is exactly identified by the linear IV approach, given that $X_i$ is uncorrelated to $e_{it}$. 

To implement our approach as in Section \ref{sec:case2}, we use third order Hermite polynomial series to approximate the nonparametric functions. Specifically, we estimate $F_{\Delta I\_{\rm US}|\Delta I\_{\rm O},X}$ at a grid of values, taken from the unconditional quantiles of $\Delta I\_{\rm US}$, using the third order Hermite polynomial basis functions of $(\Delta I\_{\rm O},X)$. Due to a large number of basis functions as the dimension of control increases, we add a ridge penalty to the least square estimation of the finite series regression, which is also a common practice in nonparametric series IV for better computation properties as suggested by \cite{newey2003instrumental}. The time fixed effects and dummy variables in $X$ are included linearly and not penalized. Using the ridge estimates, we obtain the fitted conditional CDF $\widehat{V}$ for each observation using linear interpolation. The conditional expectation $E(\Delta L|\Delta I\_{\rm US},X,V)$ is also approximated by the third order Hermite polynomials with plugged-in $\widehat{V}$, and then the average derivative estimator is obtained from the least square estimator of the series regression with ridge penalty. Both ridge penalty levels are chosen by cross-validation. For inference, we obtain confidence intervals through cluster bootstrap that resamples with replacement at the state level.  

\begin{table}[!htbp]
\centering
\caption{IV, ADE (Ridge-penalized), and Endogenous control test}
\label{tab:with_ridge}
\small
\setlength{\tabcolsep}{4pt}
\renewcommand{\arraystretch}{1.15}
\begin{threeparttable}
\resizebox{\textwidth}{!}{%
\begin{tabular}{l
S[table-format=-1.3] S[table-format=1.3] l
S[table-format=-1.3] S[table-format=1.3] l
S[table-format=-1.3] S[table-format=1.3] l}
\toprule
& \multicolumn{3}{c}{IV} & \multicolumn{3}{c}{ADE} & \multicolumn{3}{c}{IV \ -- \ ADE} \\
\cmidrule(lr){2-4}\cmidrule(lr){5-7}\cmidrule(lr){8-10}
Spec & {$\hat\beta_{p}$} & {SE} & {95\% CI}
     & {$\hat{\beta}_{np}$} & {SE} & {95\% CI}
     & {$\hat\beta_{p} - \hat\beta_{np}$} & {SE} & {95\% CI} \\
\midrule
(1) & -0.7460301 & 0.06981724 & [\num{-0.8827521}, \num{-0.6128362}] & -1.036046 & 0.1440476 & [\num{-1.2892219}, \num{-0.7383258}] & 0.2900506 & 0.1105691 & [\hspace{0.45em} \num{0.07160513}, \hspace{0.45em} \num{0.50999809}] \\
(2) & -0.6104351 & 0.10217076 & [\num{-0.7932125}, \num{-0.3987232}] & -0.49766876 & 0.08201127 & [\num{-0.679147737}, \num{-0.360950425}] & -0.1127664 & 0.1237992 & [\num{-0.3153933}, \hspace{0.45em} \num{0.1076548}] \\
(3) & -0.5376416 & 0.09367888 & [\num{-0.6869189}, \num{-0.3360142}] & -0.41961276 & 0.06889593 & [\num{-0.558874614}, \num{-0.288493628}] & -0.1180288 & 0.1145838 & [\num{-0.2618316}, \hspace{0.45em} \num{0.1110390}] \\
(4) & -0.5079711 & 0.08169406 & [\num{-0.6482154}, \num{-0.3621305}] & -0.18191397 & 0.08283669 & [\num{-0.423761791}, \num{-0.109385081}] & -0.3260547 & 0.1110428 & [\num{-0.5179489}, \num{-0.1175144}] \\
(5) & -0.5624622 & 0.08460330 & [\num{-0.7253136}, \num{-0.4054274}] & -0.12649743 & 0.07480189 & [\num{-0.285211325}, \num{-0.008018581}] & -0.4360261 & 0.1297998 & [\num{-0.6684549}, \num{-0.1838982}] \\
(6) & -0.5963601 & 0.09513239 & [\num{-0.7593581}, \num{-0.4193688}] & -0.03705099 & 0.06809177 & [\num{-0.197673461}, \hspace{0.45em} \num{0.066471710}] & -0.5594525 & 0.1271272 & [\num{-0.7798105}, \num{-0.2769113}] \\
\bottomrule
\end{tabular}%
}
\end{threeparttable}
\medskip
\smallskip

\parbox{\linewidth}{\footnotesize\textit{Note:}~The ridge penalty level is chosen by the 5-fold cross-validation with the minimum MSE rule. The bootstrap confidence interval for $\hat{\theta}$ is defined as $\mathcal{B}_n(\alpha)$ in Section \ref{sec:test}, with $\alpha= 0.05$ and 500 bootstrap replications; it is similarly defined for the other two estimators. }
\end{table}

Table \ref{tab:with_ridge} displays our results. The first IV section of the table replicates the results from Table 3 in \cite{autor2013china}, except that the standard errors and confidence intervals are replaced by the bootstrap versions to match those used for other estimators. ADE denotes the average derivative estimator with plugged-in $\widehat{V}$. The last section IV -- ADE gives the difference of these two estimators and is used for the test of endogenous controls. Six specifications in terms of controls are as follows: (1) time dummies only; (2) start-of-period employment + (1); (3) Census division specific fixed effects + (2); (4) start-of-period shares of college attainment, immigrants, female employment + (3); (5) Routine intensive occupations, offshorability index + (3); (6) All controls mentioned above. 

First, looking at the IV estimates, we find that as more controls are included, the estimated effects get smaller: the magnitude drops from 0.746 (no control), to 0.610 (one control), 0.538 (census division dummies), 0.508 (demographics), 0.562 (industrial types), and 0.596 (all controls), all with a negative sign.  The estimates of our approach display a similar pattern: the absolute sizes of the estimates reduce as more controls are included. As we discussed above, while the inclusion of more controls may solve the first type of bias, it may further introduce the second type, and our results provide alternative estimates robust to endogenous controls. If we believe more control variables help guard against the first type of bias, then the results in (6) suggest that after correcting the bias from the second type, the causal impact of Chinese import competition on the US market is not as large as what's found in the baseline results.

The standard errors of the nonparametric method are reasonably small, and some are even smaller than the linear IV estimates, which is attributed to the ridge regularization. However, the ridge penalty also introduces shrinkage bias on the slope estimators of the series regression, which in turn may cause shrinkage bias in the average derivative estimator. Therefore, we further conduct a robustness check by only penalizing the conditional CDF estimation and leaving the second-step average derivative estimator unpenalized. The results are displayed in Table \ref{tab:wo_ridge}. We find that the estimates without penalization are indeed larger than their counterparts in Table \ref{tab:with_ridge}, but the pattern remains: although the standard errors increase, we still find our approach generates smaller estimates of the average response except for the first case, where no controls are included.


\begin{table}[!htbp]
\centering
\caption{IV, ADE (un-penalized), and Endogenous control test}
\label{tab:wo_ridge}
\small
\setlength{\tabcolsep}{4pt}
\renewcommand{\arraystretch}{1.15}
\begin{threeparttable}
\resizebox{\textwidth}{!}{%
\begin{tabular}{l
S[table-format=-1.3] S[table-format=1.3] l
S[table-format=-1.3] S[table-format=1.3] l
S[table-format=-1.3] S[table-format=1.3] l}
\toprule
& \multicolumn{3}{c}{IV} & \multicolumn{3}{c}{ADE} & \multicolumn{3}{c}{IV \ -- \ ADE} \\
\cmidrule(lr){2-4}\cmidrule(lr){5-7}\cmidrule(lr){8-10}
Spec & {$\hat\beta_p$} & {SE} & {95\% CI}
     & {$\hat {\tilde {\beta}}_{np}$} & {SE} & {95\% CI}
     & {$\hat\beta_p - \hat {\tilde {\beta}}_{np}$} & {SE} & {95\% CI} \\
\midrule
(1) & -0.7460301 & 0.06981724 & [\num{-0.8827521}, \num{-0.6128362}] & -1.316694 & 0.1999447 & [\num{-1.6820738}, \num{-0.9013875}] & 0.5705717 & 0.1781184 & [\hspace{0.45em} \num{0.2370102}, \num{0.9166038}] \\
(2) & -0.6104351 & 0.10217076 & [\num{-0.7932125}, \num{-0.3987232}] & -0.6016236 & 0.1823405  & [\num{-0.97173723}, \num{-0.25080298}] & -0.008811579 & 0.2120017 & [\num{-0.3461388}, \num{0.4316790}] \\
(3) & -0.5376416 & 0.09367888 & [\num{-0.6869189}, \num{-0.3360142}] & -0.5210244 & 0.1726294  & [\num{-0.98986770}, \num{-0.29410629}] & -0.016617175 & 0.1988633 & [\num{-0.3282746}, \num{0.4702893}] \\
(4) & -0.5079711 & 0.08169406 & [\num{-0.6482154}, \num{-0.3621305}] & -0.3106442 & 0.2071021  & [\num{-0.94730909}, \num{-0.07750308}] & -0.198004944 & 0.2265179 & [\num{-0.4672718}, \num{0.4271492}] \\
(5) & -0.5624622 & 0.08460330 & [\num{-0.7253136}, \num{-0.4054274}] & -0.1382084 & 0.2120506  & [\num{-0.64956961}, \hspace{0.45em} \num{0.15101101}] & -0.424017108 & 0.2360803 & [\num{-0.7883328}, \num{0.1358488}] \\
(6) & -0.5963601 & 0.09513239 & [\num{-0.7593581}, \num{-0.4193688}] & -0.2860441 & 0.2171837  & [\num{-0.91421043}, \num{-0.06733549}] & -0.310690308 & 0.2422977 & [\num{-0.5856154}, \num{0.3486361}] \\
\bottomrule
\end{tabular}%
}
\end{threeparttable}
\medskip
\smallskip

\parbox{\linewidth}{\footnotesize\textit{Note:}~The ridge penalty level (for the conditional CDF estimation) is chosen by the 5-fold cross-validation with the minimum MSE rule. The bootstrap confidence interval for $\hat{\theta}$ is defined as $\mathcal{B}_n(\alpha)$ in Section \ref{sec:test}, with $\alpha= 0.05$ and 500 bootstrap replications; it is similarly defined for the other two estimators. }
\end{table}

Lastly, the endogenous control test is conducted through estimates $\hat\beta_{p} -\hat\beta_{np} $ and the bootstrap 95\% confidence interval from the last sections of Tables \ref{tab:with_ridge} and \ref{tab:wo_ridge}. Since row (1) does not include a control variable, our test does not apply there. Examining rows (2) - (6) in Table \ref{tab:with_ridge}, we find rejections of no endogenous control in the last three rows, suggesting the ADEs are preferable in specifications (4) - (6), where the 95\% confidence interval excludes zero. Due to larger standard errors, the test results in Table \ref{tab:wo_ridge} are not very informative, given that none of the tests are rejected.

\section{ Conclusion}
\label{sec:conclu}

We address a critical, prevalent, yet often overlooked problem in empirical research: the endogeneity of control variables. Building on the insightful observation and discussion in \cite{frolich2008parametric} that nonparametric estimation can help with the endogenous control problem, we provide constructive identification results for structural average response in a broad class of models. The results show that the control endogeneity needs not prevent identification when conditional validity and sufficient conditional treatment variation hold. 

For empirical studies, our results invite researchers to conduct robustness checks using nonparametric methods and the proposed test. The rejection of the specification test should be interpreted as a robustness warning, with an endogenous control interpretation only under the maintained structural model.

\singlespacing

\section*{\centering  Data Availability Statement}
The China shock data are available from the public replication archive of \cite{autor2013china}. Replication code and data documentation are included with the submission; file placement and software requirements are described in the replication README.

\section*{\centering Disclosure Statement}
The authors report there are no competing interests to declare.

\bibliographystyle{apalike}
\bibliography{endo_controls.bib}

\appendix
\section*{Appendix}

\begin{proof}[Proof of Theorem \ref{thm1}]
  By Assumptions \ref{assum:cond_ind}, $f_{\varepsilon|W=w,\widetilde{X}=\tilde{x}}(\epsilon)= f_{\varepsilon|\widetilde{X}=\tilde{x}}(\epsilon)$. Then, applying Leibniz integral rule under the existence of an integrable dominating function and the chain rule gives
\begin{align*}
    &\partial _w E[Y|W=w,\widetilde{X} = \tilde{x}] = \partial_w \int m(w,x,\epsilon)f_{\varepsilon|W=w,\widetilde{X} = \tilde{x}}(\epsilon) d\epsilon \\
    =& \int \partial_w m(w,x,\epsilon) f_{\varepsilon|W=w,\widetilde{X} = \tilde{x}}(\epsilon)d\epsilon = E[\partial_w m(w,x,\varepsilon)|W=w,\widetilde{X} = \tilde{x}] 
\end{align*}
Case 1: If $\widetilde{X} = X$, then $\partial _w E[Y|W=w,\widetilde{X} = \tilde{x}]= \beta(w,x)$, and so $\beta =  \int \beta(w,x)d F(w,x) = E\left[\partial _W E\left[Y|W,\widetilde{X}\right]\right]$. Case 2: If $\{\widetilde{X}\backslash X\} \ne \emptyset$, by integrating out $\{\widetilde{X}\backslash X\}$, 
\begin{align*}
   E[ \partial _w E[Y|W=w,\widetilde{X} = \tilde{x}] |W=w,X=x] = E[\partial_w m(w,x,\varepsilon)|W=w,X = x] =\beta(w,x) 
\end{align*}
Then, by integrating out $W$ and $X$, $\beta =  \int \beta(w,x)d F(w,x) = E\left[\partial _W E\left[Y|W,\widetilde{X}\right]\right]$. Assumption~\ref{assum:local_support} ensure the derivative of the conditional mean is identified from the observed conditional distribution.

When $Y$ is binary, given $\varepsilon$ is partitioned as $(u,v)$,
\begin{align*}
    &E[Y|W=w,\widetilde{X} = \tilde{x}] = \int 1\{m^*(w,x,e)>0 \}f_{\varepsilon|W=w,\widetilde{X} = \tilde{x}}(e)de \\
    = & \int \int 1\{m^*(w,x,u,v)>0 \}f_{u,v|W=w,\widetilde{X} = \tilde{x}}(u,v)dudv 
    = \int \left[\int_{u>u^*(w,x,v)}f_{u,v|W=w,\widetilde{X} = \tilde{x}}(u,v)du\right]dv.
\end{align*}
where the last equality follows because $m^*(w,x,u,v)$ strictly increases in $u$ and $u^*(w,x,v)$ is the unique solution of  $m^*(w,x,u,v)=0$. By implicit function theorem under the conditions in the statement, $u^*(w,x,v)$ is differentiable in $w$. Then, by Leibniz rule, chain rule, CIA, and the measurable separability between $W$ and $\widetilde{X}$, we have
\begin{align*}
    \partial_wE[Y|W=w,\widetilde{X} = \tilde{x}]  = \int  - \partial_w u^*(w,x,v) f_{u,v|W=w,\widetilde{X} = \tilde{x}}(u^*(w,x,v),v) dv .
\end{align*}
Then, similarly, by considering Case (1) $\widetilde{X} = X$ and Case (2) $\{\widetilde{X}\backslash X\} \ne \emptyset$ as above, we have $\beta^*(w,x) = \partial _w E[Y|W=w,\widetilde{X}=\tilde{x}]$ under Case (1) and $\beta^*(w,x) = E[ \partial _w E[Y|W=w,\widetilde{X} = \tilde{x}] |W=w,X=x]$ under Case (2). Therefore, by integrating out $W$ and $X$, we conclude that $\beta^* = \int \beta^*(w,x)d F(w,x) = E\left[\partial _W E\left[Y|W,\widetilde{X}\right]\right]$, which completes the proof.
\end{proof}

\begin{proof}[Proof of Theorem \ref{thm2}]

For statement (i), let $l$ be any continuous and bounded real function. Using the independence of $Z$ and $\varepsilon$ conditional on $X$, we first obtain the conditional mean independence as an intermediate result:
\begin{align*}
    E[l(W)|\varepsilon,\eta,X]  = \int l(q(z,X,\eta))dF_{Z|\varepsilon,\eta,X}(z) 
      = \int l(q(z,X,\eta))dF_{Z|\eta,X}(z) = E[l(W)|\eta,X].
\end{align*}
Then, we can check CIA of $W$ and $\varepsilon$ given $(\eta,X)$ by a conditional version of Theorem 2.1.12 of \cite{durrett2019probability}. Let $a(\cdot)$ and $b(\cdot)$ be any continuous and bounded real functions, then
\begin{align*}
    E[a(W)b(\varepsilon)|\eta,X] & = E[E[a(W)b(\varepsilon)|\varepsilon,\eta,X]|\eta,X]  = E[E[a(W)|\varepsilon,\eta,X] b(\varepsilon)|\eta,X] \\
    & =  E[E[a(W)|\eta,X] b(\varepsilon)|\eta,X]  = E[a(W)|\eta,X]E[b(\varepsilon)|\eta,X].
\end{align*}

Consider statement (ii). Under Assumption \ref{assum:v_eta}(i), there exists an inverse function $q^{-1}(Z,X,.): \mathcal{W}\to \text{supp}(\eta)$ such that $q^{-1}(Z,X,q(Z,W,\eta)) = \eta$, a.s. Then, we have
\begin{align*}
   F_{W|Z,X}(w) & = Pr(W\leq w|Z,X) = Pr(q(Z,X,\eta)\leq w|Z,X) \\
     &= Pr(\eta\leq q^{-1}(Z,X,w)|Z,X) = F_{\eta|X}(q^{-1}(Z,X,w)).
\end{align*}
where the last equality follows from the independence of $Z$ and $\eta$ conditional on $X$. Note that $\eta = q^{-1}(Z,X,W)$ a.s., so we have
$V = F_{W|Z,X}(W) = F_{\eta|X}(\eta)$. Let $Q_{\eta|X}(u) = \inf\{a\in \mathbb{R}:F_{\eta|X}(a)\geq u \}$ be the conditional quantile function of $\eta$ given $X$. Under Assumption \ref{assum:v_eta}(ii), $Q_{\eta|X}(F_{\eta|X}(\eta)) = \eta$. We note that $\sigma(F_{\eta|X}(\eta),X) \subset \sigma(\eta, X)$ since $F_{\eta|X}(\eta)$ is a function of $\eta$ and $X$. Because $(\eta,X) = (Q_{\eta|X}(F_{\eta|X}(\eta)),X) $, we also have $\sigma(\eta,X)\subset \sigma(F_{\eta|X}(\eta),X)$. By setting $V= F_{\eta|X}(\eta)$, it follows that 
\begin{align*}
    E[a(W)b(\varepsilon)|V,X] & = E[a(W)b(\varepsilon)|\eta,X] \\
    & = E[a(W)|\eta,X]E[b(\varepsilon)|\eta,X] = E[a(W)|V,X]E[b(\varepsilon)|V,X]
\end{align*}
which implies condition \eqref{eq:controlv}.    
\end{proof}

\begin{proof}[Proof of Theorem \ref{thm:3}]
Under the conditions of Theorem \ref{thm1}, regardless of whether $X$ are endogenous or exogenous, $\beta_0  = E[\partial_W g(W,X)] =\beta_{np}$. By Assumption \ref{assum:test} (i), $\theta_0= 0$ under the null and $\theta_0 = B_n$ under the alternative. Then, under the null, by \eqref{eq:asym_linear_p} and \eqref{eq:asym_linear_np} as well as Assumption \ref{assum:test} (iv), 
\begin{align*}
    \sqrt{n}\hat{{\theta}} =&   \sqrt{n}\left(\hat{\beta}_{p} - \beta_{p}\right) +   \sqrt{n}\left(\beta_{np} - \hat\beta_{np}\right)  \\
    =& \frac{1}{  \sqrt{n}}\sum_{i=1}^n\psi (W_i,X_i) + o_P(1) \overset{d}{\to} N(0,V)
\end{align*}
where the last equality follows from Lindeberg–Lévy CLT and $V=\text{Var}(\psi(W,X))$. 

Consider the bootstrap root $\sqrt{n}\left(\hat{\theta}^*-{\hat\theta}\right)$. By \eqref{eq:asym_linear_p_boot} and \eqref{eq:asym_linear_np_boot},
\begin{align*}
    \sqrt{n}\left(\hat{\theta}^*-{\hat\theta}\right) = \frac{1}{\sqrt{n}}\sum_{i=1}^n \psi(W_i^*,X_i^*)- \psi(W_i,X_i) + o_{P^*}(n^{-1/2})\overset{d^*}{\to} N(0,V)
\end{align*}
where the weak convergence conditional on the sample $\{Y_i,W_i,X_i\}_{i=1}^n$ follows from Theorem 10.4 in \cite{hansen2022econometrics}. By Lemma 23.3 in \cite{Vaart_1998}, with the standard errors replaced by constant 1, $ P\{0 \in \mathcal{B}_n(\alpha) | H_0\} \to  1-\alpha$.

Under the alternative, we have
\begin{align*}
 \sqrt{n} \left( \hat{{\theta}} - B_n \right) = \frac{1}{  \sqrt{n}}\sum_{i=1}^n\psi (W_i,X_i) + o_P(1) \overset{d}{\to} N(0,V)
\end{align*}
Let $\mathcal{B}_n(\alpha) - B_n = \{\theta - B_n: \theta\in \mathcal{B}_n(\alpha)\}$. Then, by Lemma 23.3 in \cite{Vaart_1998} again, $ P\{0 \in \mathcal{B}_n(\alpha) - B_n| H_1\} \to  1-\alpha$. It follows that $ P\{ B_n \in \mathcal{B}_n(\alpha)| H_1\} \to  1-\alpha$. Due to asymptotic normality of $ \sqrt{n}\left(\hat{\theta}^*-{\hat\theta}\right)$, the confidence region shrinks at the rate $n^{-1/2}$ as $n\to \infty$. Since $\sqrt{n}|B_n|\to\infty$, $|B_n|$ dominates the (half) length of the confidence interval, thus $P\{ 0 \notin \mathcal{B}_n(\alpha)|H_1\} \to 1$, which completes the proof.
\end{proof}

\end{document}